\documentclass{ws-ijmpcs_modified}

\usepackage{graphics}
\usepackage{amssymb}
\newcommand\als{\alpha_{\rm s}}
\newcommand\lQ{\Lambda_{QCD}}
\newcommand{\be}{\begin{equation}}
\newcommand{\ee}{\end{equation}}
\newcommand{\bea}{\begin{eqnarray}}
\newcommand{\eea}{\end{eqnarray}}

\begin{document}

\markboth{Joan Soto}
{Overview of charmonium decays and production from Non-Relativistic QCD}

%
%

\title{\hfill ICCUB-11-002\\ \hfill UB-ECM-PF 10/44\\ $\;$\\$\;$\\OVERVIEW OF CHARMONIUM DECAYS AND PRODUCTION FROM NON-RELATIVISTIC QCD
}
\author{JOAN SOTO
}

\address{Departament d'Estructura i Constituents de la Mat\`eria and Institut de Ci\`encies del Cosmos,
Universitat de Barcelona, Mart\'\i $\;$ i Franqu\`es 1\\
08028 Barcelona, Catalonia, Spain
\\
joan.soto@ub.edu}

\maketitle

\begin{history}
\end{history}

\begin{abstract}
I briefly review Non-Relativistic QCD and related effective theories, and discuss applications to 
heavy quarkonium decay, and production in electron-positron colliders. 

\end{abstract}

\ccode{PACS numbers: 12.38.-t, 14.40.Pq, 12.39.St, 13.20.Gd, 13.25.Gv}

\section{Introduction}
\label{intro}
Heavy quarkonia are mesons made out of a heavy quark and a heavy antiquark 
whose masses ($m_Q$) are larger 
than $\lQ$, the typical hadronic scale\cite{Brambilla:2004wf}.
 These include bottomonia ($b\bar b$), charmonia ($c \bar c$), $B_c$ systems ($b \bar c$ and $c \bar b$)
 and would-be toponia ($t\bar t$). 
In the quarkonium rest frame the heavy quarks move slowly ($v \ll 1$, $v$ being the typical heavy quark velocity in the center of mass frame), 
with a typical momentum $m_Qv \ll m_Q$
and binding energy $\sim m_Qv^2$. Hence any study of heavy quarkonium faces a multiscale problem with the hierarchies $m_Q \gg m_Qv \gg m_Qv^2$ 
and $m_Q \gg \lQ$. The use of effective field theories is extremely convenient in order to exploit these hierarchies.
 Based on the pioneering work of Caswell and Lepage\cite{Caswell:1985ui}, a systematic approach to study 
these systems from QCD has been 
developed 
which is generically known as Non-Relativistic QCD (NRQCD)\cite{Bodwin:1994jh}. 
The NRQCD formalism for spectroscopy and
 inclusive decays to light particles is very well understood (this is also so for electromagnetic threshold production)
so that NRQCD results may be considered QCD results up to a given order in the expansion parameters\cite{Brambilla:2004jw}
 (usually $\als (m_Q)$ and $1/m_Q$).
The NRQCD formalism for production is more controversial. I will summarize here the main 
features of NRQCD (and related effective theories) and review recent theoretical progress in decays, and production at electron-positron colliders.
I refer to the contribution of Geoff Bodwin for an overview on production at electron-proton and hadron colliders,
 and on the status of factorization 
proofs\cite{Bodwin:2010py}.
 
\section{Non-Relativistic QCD}
\label{NRQCD}
The hierarchy of scales exploited in NRQCD is 
$m_Q \gg m_Qv , m_Qv^2 , \lQ$.
The part of the NRQCD Lagrangian bilinear on the heavy quark fields coincides with the one of Heavy Quark Effective Theory
 (HQET)\cite{Neubert:1993mb}, but the size assigned to each term differs from it.
It also contains analogous bilinear terms for the antiquark, and four fermion operators\cite{Bodwin:1994jh}, 
which are classified as color singlet or color octet. 
The short distance matching coefficients 
of these operators have imaginary parts.
The fact that NRQCD is equivalent to QCD at any desired order in $1/m_Q$ and $\als (m_Q)$ makes the lack of unitarity innocuous. 
In fact, it is turned into an advantage: it facilitates the calculation of inclusive decay rates to light particles.

The inclusive decay widths of heavy quarkonium states to light particles are given in terms of sums of NRQCD matrix 
elements of both color singlet and color octet operators multiplied by the imaginary part of short distance matching
 coefficients, which can be calculated in perturbation
 theory in $\als (m_Q)$. 
Once a size in terms of $m_Q$ and $v$ is assigned to each of the matrix elements, a systematic expansion in powers of 
$v$ and $\als (m_Q)$ is obtained. 
 Earlier QCD factorization formulas 
were missing the matrix elements of color octet operators. They are inconsistent because the color octet matrix elements 
are necessary to cancel the 
factorization scale dependence which arises in loop calculations of the short distance matching 
coefficients\cite{Bodwin:1992ye}. 
The matrix elements cannot be calculated in perturbation theory of $\als (m_Q)$. The color singlet ones are related to wave functions at the origin 
but the color octet ones are not, in the general case. Sometimes only the effect of color octet operators in the 
renormalization group
evolution equations is taken into account\cite{Bodwin:1994jh,Fan:2009cj}. If $m_Q v\gg \lQ$ they can be estimated using weak coupling techniques 
\cite{GarciaiTormo:2004jw,GarciaiTormo:2007qs} and if $m_Q v^2\ll \lQ$ they can be related to wave functions at the origin plus a number of universal 
non-perturbative parameters\cite{Brambilla:2001xy}.
They can always be calculated on the lattice\cite{Bodwin:2001mk}. Alternatively they can also be extracted from 
data\cite{Maltoni:2000km}. The imaginary parts of the matching coefficients of the dimension 6 and dimension 8 
operators are 
known 
at ${\cal O}(\als^3)$\cite{Petrelli:1997ge}, those of 
dimension 10 ($S$ and $P$ waves) at ${\cal O}(\als^2)$\cite{Brambilla:2006ph,Bodwin:2002hg,Ma:2002ev,Huang:1997nt,Brambilla:2008zg}, 
those of dimension 10 ($D$ wave) at ${\cal O}(\als^3)$\cite{He:2008xb,Fan:2009cj,He:2009bf}, and 
the one of the color singlet $^{3}S_1$ operator at ${\cal O}(\als^4)$
\cite{Mackenzie:1981sf,Campbell:2007ws}.

\section{Potential NRQCD}

Unlike HQET, 
the NRQCD Lagrangian does not enjoy a homogeneous counting, due to the fact that the scales $m_Qv$, $m_Qv^2$ 
and $\lQ$ are still entangled. 
Potential NRQCD (pNRQCD) aims at disentangling these scales, and hence to facilitate the counting, by constructing 
a further effective theory in which energy scales larger $m_Q v^2$ are integrated out\cite{Pineda:1997bj}. 
If $\lQ \lesssim m_Q v^2$, then $m_Q v \gg \lQ$ and the matching between NRQCD and pNRQCD can be carried out 
in perturbation theory in $\als (m_Qv)$ . This is the so called weak coupling regime. If $\lQ \gg m_Q v^2$, the matching cannot be carried out 
in perturbation theory in $\als (m_Qv)$ anymore, but one can still exploit the hierarchy $m_Q \gg m_Q v , \lQ \gg m_Q v^2$. This is the so called 
strong coupling regime.
Since $m_Q$, $v$ and $\lQ$ are not directly observable, given heavy quarkonium state it is not clear to which of the above regimes, 
if to any\footnote{States close or above the open flavor threshold are expected to belong neither the weak nor the strong coupling regimes.} 
, it must be assigned to. A test was proposed\cite{GarciaiTormo:2005bs} using the photon spectra of radiative decays. Under this test,
 CLEO data\cite{Besson:2005jv}
suggests that $\Upsilon (2S)$ and $\Upsilon (3S)$ are in the strong coupling regime whereas $\Upsilon (1S)$ is not. An alternative test 
using leptonic decays was also carried out 
leading to the same conclusion\cite{DomenechGarret:2008vk}. Even though experimental data is now available\cite{Besson:2008pr,Libby:2009qb}
 and it was suggested at some 
point\cite{GarciaiTormo:2007qs}, no such a test for charmonium states 
has been carried out yet.  

\subsection{Weak Coupling Regime}

The pNRQCD Lagrangian in this regime can be written in terms of
a color singlet and a color octet heavy quark-antiquark wave function fields that evolve according to 
the Hamiltonians $h_s$ and $h_o$ respectively, and interact with gluons of energy $\sim m_Q v^2$ (ultrasoft). The potentials in those Hamiltonians,
 as well as the remaining
 matching coefficients, may be obtained by matching to NRQCD in perturbation theory in $\als (m_Qv)$ and $1/m_Q$ 
at any order of the multipole expansion ($
1/m_Qv$). The static potential in $h_s$ 
is known up to  
three loops\cite{Peter:1997me,Schroder:1998vy,Brambilla:1999qa,Brambilla:1999xf,Anzai:2009tm,Smirnov:2009fh} 
(the logarithmic contributions at four loops  are also known\cite{Brambilla:2006wp}), and in $h_o$ up to two loops\cite{Kniehl:2004rk}. 
The renormalon singularities in the static potentials are also understood in some 
detail\cite{Beneke:1998rk}. 
The $1/m_Q$ and $1/m_Q^2$ terms in $h_s$ are known at two and one loop respectively\cite{Kniehl:2002br}. 
This Lagrangian has been used to carry our calculations at fixed order in $\als$: a complete NNNLO 
(assuming $\lQ \ll m_Q\als^2$) expression
 for the spectrum is available 
\cite{Penin:2002zv,Smirnov:2009fh}. Most remarkably resummations of logarithms can also be carried out using renormalization group 
techniques\cite{Pineda:2000gza,Pineda:2001ra,Brambilla:2009bi}.
Thus the hyperfine splitting and the ratio vector/pseudoscalar electromagnetic decay width have been calculated at 
NNLL and NLL respectively\cite{Kniehl:2003ap,Penin:2004ay}. The hyperfine splitting compares well to data for 
charmonium, but undershoots 
considerably the experimental value 
for bottomonium. A reorganization of the quantum mechanical calculation has been recently proposed that may resolve this 
problem\cite{Kiyo:2010zz}.
 At least it brings the ratio of the vector/pseudoscalar electromagnetic decays mention above in good agreement with 
experiment for charmonium.

\subsection{Strong Coupling Regime}

The pNRQCD Lagrangian in this regime reduces to a heavy quark-antiquark wave function field interacting with a 
potential\footnote{We ignore for simplicity pseudo-Goldstone bosons.}, which can be organized in powers of $1/m_Q$. These potentials  
 cannot be calculated in perturbation theory of $\als (m_Qv)$ anymore, but can 
be calculated in lattice simulations\cite{Bali:1997am}.
They must coincide at short distances with perturbative
 calculations\cite{Brambilla:2010pp} and at long distances they must be compatible with the effective string theory of 
QCD\cite{PerezNadal:2008vm}.
The fact that $\lQ \gg m_Qv^2$ can now be exploited to further factorize NRQCD decay matrix elements 
 into wave functions at the origin and universal bound state independent 
parameters\cite{Brambilla:2001xy,Brambilla:2003mu}.
 This further factorization allows to put forward new model independent predictions, for instance the ratios of hadronic 
decay widths of P-wave 
states in bottomonium were predicted from charmonium data\cite{Brambilla:2001xy}, or the ratio of photon spectra in 
radiative decays of vector 
resonances\cite{GarciaiTormo:2005bs}.

\section{Beyond Inclusive Decays}

NRQCD can in principle describe transitions between heavy quarkonium states, 
but no useful information has been extracted so far from it, beyond the fact that non-relativistic hadronic effective 
theories implementing heavy quark and chiral symmetries\cite{Casalbuoni:1996pg,DeFazio:2008xq,He:2010pb},
may be considered hadronic realizations of it. 
pNRQCD can also describe transitions between heavy quarkonium states below open flavor threshold, and 
has been applied to magnetic dipole transitions\cite{Brambilla:2005zw}.
Decays to heavy light meson pairs can in principle be described by NRQCD as well, but again no much information has 
been obtained from it so far. 

For exclusive decays and for certain kinematical end-points of semi-inclusive decays to light particles, NRQCD must be 
supplemented with collinear degrees of freedom. This can be done in the effective theory framework of 
SCET\cite{Bauer:2000ew}. Exclusive radiative decays of heavy quarkonium in SCET have been 
addressed\cite{Fleming:2004hc}, where results analogous to those of traditional light cone factorization formulas have
 been obtained\cite{Ma:2001tt}. Recently, SCET, HQET, NRQCD and pNRQCD have been combined to obtain model independent 
formulas for bottomonium decays to $D\bar D$ meson pairs\cite{Azevedo:2009dt}.
Concerning semi-inclusive decays, a remarkable success has been achieved by combining  SCET, NRQCD and pNRQCD for the 
photon spectrum of $\Upsilon (1S) \rightarrow \gamma X$,
a challenge since the early days of 
QCD\cite{Fleming:2002rv,GarciaiTormo:2004jw,GarciaiTormo:2005ch}. Unfortunately, when this formalism is applied to 
radiative decays of $J/\psi$ the outcome suffers from large uncertainties\cite{GarciaiTormo:2007qs,GarciaiTormo:2009zz}.

\section{Production}

In production processes, like in semi-inclusive and exclusive decays, gluons of energies $\sim m_Q$ may exists in the initial or final states, 
and hence NRQCD must be supplemented with additional factorization formulas in order to disentangle them from the dynamics of heavy quarkonium.

\subsection{Electromagnetic threshold production}

This is the simplest and best understood production process because no gluons of energies $\sim m_Q$ exist in 
the initial and final states. In the weak coupling regime, which is relevant for a 
precise measurement of top quark mass in the future ILC, the cross-section at NNLO is known for some 
time\cite{Hoang:2000yr} and the log resummation at NLL is also available\cite{Pineda:2001ra,Hoang:2002yy}. 
The calculations at NNNLO are almost complete\cite{Kniehl:1999mx}, and partial results also 
exist for the log resummation at NNLL\cite{Hoang:2002yy,Pineda:2006ri}. At this level of precision electroweak 
effects must also be taken into account in the $t\bar t$ case\cite{Hoang:2004tg}. 

\subsection{Inclusive production}

A factorization formula for the inclusive production of heavy quarkonium was put forward in the framework of 
NRQCD\cite{Bodwin:1994jh}, which was assumed to hold provided 
that the transverse momentum ${\bf p_\perp}$ was larger or of the order of the heavy quark mass.   
The formula contains a partonic level cross section, which can be calculated in perturbation theory in $\als (m_Q)$,
in which the heavy quark pair may be produced in color singlet or a color octet state, and long distance production 
matrix elements 
which encode the evolution of the pair to the actual physical state, which cannot be calculated in perturbation theory 
in $\als (m_Q)$, and are
usually extracted from data. 
The production matrix elements were
assigned sizes according to the NRQCD velocity scaling rules\cite{Bodwin:1994jh}, which correspond to 
the weak coupling regime. 
This factorization formalism is receiving a closer look in the framework of fragmentation 
functions\cite{Nayak:2007mb} (${\bf p_\perp} \gg m_Q$),
and has been proved to be correct at NNLO in $\als (m_Q)$, provided a slight redefinition of the matrix elements 
is carried out\cite{Nayak:2005rw}. 

The measurement of $\sigma (J/\psi + c\bar c+X)/\sigma (J/\psi+X)$ at Belle
in 2002\cite{Abe:2002rb} 
was in clear conflict with LO NRQCD 
results\cite{Cho:1996cg}, and has triggered a number higher order calculations. 
For $\sigma (J/\psi + c\bar c+X)$, a NLO 
calculation of the color singlet contribution
in $\als (m_Q)$ gives a large correction\cite{Zhang:2006ay,Gong:2009ng}. This is also the case for the NLO color octet
 contribution\cite{Liu:2003jj,Zhang:2009ym}.
Relativistic corrections\cite{He:2007te} and two-photon
 contributions\cite{Liu:2003zr} have also been calculated. For $\sigma (J/\psi+X({\rm non-}c\bar c))$, the LO color octet\cite{Wang:2003fw} and the NLO color single both in  
$\als (m_Q)$\cite{Ma:2008gq,Gong:2009kp}
 and $v$\cite{He:2009uf} contributions have been calculated. When all these results are put together there is no 
obvious 
discrepancy with experiment anymore, but the theoretical uncertainties are still large.

Near certain kinematical end-points NRQCD production processes must be supplemented with collinear degrees of freedom, 
in analogy to semi-inclusive decays discussed above. This issue has been addressed using 
SCET\cite{Fleming:2003gt}.

\subsection{Exclusive production}

The basic ideas of NRQCD factorization have also been applied 
to exclusive processes, mostly after the surprisingly large double charmonium cross-section first measured at 
Belle\cite{Abe:2002rb}. Recently, factorization proofs have been put forward for this process as well as for exclusive 
production in 
$B$-decays\cite{Bodwin:2008nf}. Although Belle and Babar results for $J/\psi + \eta_c$ 
production cross section 
do not quite agree with each other, both numbers are roughly an order of magnitude larger than LO NRQCD 
predictions\cite{Liu:2002wq}. A NLO calculation in $\alpha (m_Q)$ gives a very large 
correction\cite{Zhang:2005cha,Gong:2007db} which, together with a number of relativistic 
corrections\cite{Bodwin:2006dn},
brings the theoretical calculation in agreement with both experimental results within errors\cite{He:2007te,Bodwin:2007ga}.

\section*{Acknowledgments}
I am grateful to Geoff Bodwin and Zhiguo He for assistance in confectioning the talk. I acknowledge financial 
support from the ECRI HadronPhysics2
(Grant Agreement n. 227431) (EU), the CSD2007-00042 Consolider-Ingenio 2010 program, the
FPA2007-66665-C02-01/, 
 FPA2007-60275/MEC and FPA2010-16963 grants
(Spain), and the 2009SGR502 CUR grant (Catalonia).

\end{document}